\documentclass[12pt]{iopart}

\usepackage{iopams}  
\usepackage{graphicx}  
\usepackage{epsf}
\begin{document}

\title[New method to integrate (2+1)-wave equations with Dirac's delta functions]
{A new method to integrate (2+1)-wave equations with Dirac's delta functions as sources}

\author{Carlos O. Lousto and Hiroyuki Nakano}

\address{Center for Computational Relativity and Gravitation, 
School of Mathematical Sciences, 
Rochester Institute of Technology, Rochester, New York 14623, USA}
\ead{colsma@rit.edu, hxnsma@rit.edu}
\begin{abstract}
Unlike in the Schwarzschild black hole background, 
gravitational perturbations in a Kerr black hole background 
can not be decomposed into simple tensor harmonics in the time domain. 
Here, we make mode decompositions only in the azimuthal direction. 
As a first step, we discuss the resulting (2+1)-dimensional 
Klein-Gordon differential equation 
for scalar perturbations with a two dimensional Dirac's $\delta$-function 
as a source representing a point particle orbiting a much larger black hole. 
To make this equation amenable for numerical integrations 
we explicitly remove analytically the singular behavior of the source 
and compute a global, well behaved, effective source for the corresponding waveform. 

\end{abstract}

\submitto{\CQG}
\maketitle

%%%%%%%%%%%%%%%%%%%%%%%%%%%%%%%%%%%%%%%%%%%%%%%%%%%%%%%%%%%%%%%%%%%%%%
\section{Introduction}
%%%%%%%%%%%%%%%%%%%%%%%%%%%%%%%%%%%%%%%%%%%%%%%%%%%%%%%%%%%%%%%%%%%%%%

One of the main astrophysical targets of LISA \cite{LISA},
a space-based interferometric gravitational wave detector, 
is the gravitational waves 
generated by the inspiral of compact objects into massive black holes. 
To extract physical information of such 
extreme mass ratio inspirals (EMRI), 
it is important to know the theoretical gravitational waveforms 
with sufficient accuracy. 
For the EMRI scenario, we use the black hole perturbation approach 
to compute waveforms. Here the compact object is approximated 
by a point particle orbiting a massive Kerr black hole. 
In order to obtain the precise theoretical gravitational waveforms 
we need to solve the self-force problem 
\cite{Poisson:2004gg,Hikida:2004hs,CQGVol22No15} 
and then for second order perturbations 
\cite{Nakano:2007hv,CQGVol22No15}. 

It has already been over two years 
that numerical relativity produced one 
of the most spectacular breakthroughs 
in science \cite{Pretorius:2005gq,Campanelli:2005dd, Baker:2005vv},
 succeeding in solving 
the two body problem in general relativity after decades of effort. 
Among the notable set of results, 
the discovery \cite{Campanelli:2007ew} of large recoil velocities 
(up to $4000\,km/s$) \cite{Campanelli:2007cga} stands out. 
Notably in Ref.~\cite{Campanelli:2007ew} 
a generic binary black hole case was treated, with unequal spins and 
unequal masses (mass ratio 1/2). 
It has been hard to deal with extreme mass ratios 
and recent computations limit to $m_1/m_2=3/8$ \cite{Lousto:2007db} 
for spinning holes 
and up to nearly 1/4 for nonspinning holes \cite{Gonzalez:2006md}. 
It is foreseeable that soon mass ratios of nearly 1/10 can be 
achieved in this full numerical simulations making possible a 
comparison with the semi-analytic approach of the self-force. 

The self force problem was formally resolved over ten years ago 
\cite{Mino:1996nk,Quinn:1996am}, 
but its implementation in explicit computations proved lengthy 
and difficult. This have been reformulated in a more elaborated way 
by Detweiler and Whiting \cite{Detweiler:2002mi}. 
The first corrections to the trajectory of an EMRI was 
computed in \cite{Lousto99b} for a headon collision 
using the Regge-Wheeler gauge and the $\zeta$-function regularization. 
Those results were later confirmed 
using the standard formalism \cite{Barack:2001gx,Mino:2001mq,Barack:2002bt} 
in the Lorenz gauge \cite{Barack:2002ku}. 
In order to compute the generic orbit corrections around a Schwarzschild 
black hole, Barack and Lousto \cite{Barack:2005nr} have 
approached the problem of directly solving the linearized 
Einstein equations in the Lorenz gauge 
instead of the Regge-Wheeler and Zerilli wave equations \cite{Nakano:2003he}. 
To do that, one needs to be able to develop an accurate algorithm 
to integrate ten coupled wave-like equations with sources proportional 
to a Dirac's delta. This algorithm has been developed 
\cite{Lousto97b,Lousto:2005qh} for the (1+1)-wave equation 
resulting from the tensor harmonic decomposition of perturbations 
in the Schwarzschild background, 
but not for the (2+1)-equation resulting from perturbations of a 
Kerr (spinning) black hole. This is the subject of the current work. 

In this paper, we focus on one aspect of the self-force problem, 
specifically to derive the full, bare or retarded field of a point source. 
Therefore, we do not treat the local analysis of the field 
around the particle's location because the retarded field is global. 
Here, instead of studying the Teukolsky equation 
for the curvature perturbations 
$\psi_4$, as a first step, we consider the Klein-Gordon equation 
in the Schwarzschild spacetime, but do not decompose it 
into spherical harmonics, in order to model perturbations 
like in the more generic Kerr background. 
Recently, introducing a thin worldtube surrounding the worldline 
of a point particle, Barack and Golbourn \cite{Barack:2007jh} 
have discussed this equation in (2+1)-dimensions as 
derived by the mode decomposition in the azimuthal direction. 
A different treatment is proposed here to deal with this problem globally. 
There is also a method to approximate a Dirac's $\delta$-function 
by narrow Gaussian \cite{LopezAleman:2003ik}. 
It is, however, difficult to ascertain the error introduced 
by smearing the particle and if this is accurate enough 
for self force computations. 

Once we obtain the retarded field, 
each azimuthal mode of the self-force on the particle 
can be calculated and is finite at the particle location. 
But the summation over all azimuthal modes diverges. 
Therefore, we need some regularization 
to derive the regularized self-force. 
At this stage, it is necessary to discuss the local analysis 
of the field or self-force in the derivation of the singular part. 
The regularized self-force includes two parts, i.e., 
the conservative part and the dissipative part \cite{Pound:2007th}. 
To obtain the conservative part of the self-force, we need 
the regularization, while it is not necessary 
for the dissipative part 
which is derived by using a radiative Green's function \cite{Mino:2003yg}. 
Recently, Barack, Golbourn and Sago formulated 
a new scheme to construct the regularized self force directly 
from the azimuthal modes of the field in \cite{Barack:2007we}. 

The paper is organized as follows. 
In \sref{sec:form}, we discuss 
the (2+1)-dimensional Klein-Gordon differential equation 
with a 2-dimensional $\delta$-function as a source. 
To remove the $\delta$-function, we introduce a new wave-function. 
This formulation is done in the case of general orbits 
in the Schwarzschild background. 
In \sref{sec:COC}, we apply the formulation given in \sref{sec:form} 
to the case of circular orbits. Here, we obtain a global effective source 
which is well behaved everywhere. To do so, 
we also treat boundary behaviors both near the black hole horizon 
and at spatial infinity. 
In \sref{sec:dis}, we summarize the results of this paper 
and discuss its applications. 
Some details of the calculations are given in the appendices.
Throughout this paper, we use units in which $c=G=1$.

%%%%%%%%%%%%%%%%%%%%%%%%%%%%%%%%%%%%%%%%%%%%%%%%%%%%%%%%%%%%%%%%%%%%%%
\section{Formulation} \label{sec:form}
%%%%%%%%%%%%%%%%%%%%%%%%%%%%%%%%%%%%%%%%%%%%%%%%%%%%%%%%%%%%%%%%%%%%%%

When we calculate the (2+1)-dimensional equation 
derived from the 4-dimensional Klein-Gordon equation 
by the azimuthal mode decomposition, 
the resulting equation is not exactly same 
as the (2+1)-dimensional wave equation. 
In our formulation, it is important 
to derive a differential operator 
which is the (2+1)-dimensional d'Alambertian 
of the flat spacetime. 
By transforming the scalar field, we can obtain an equation 
which includes the flat (2+1)-dimensional d'Alambertian and a remainder
as in \eref{eq:formal} below. 
Then, we remove the 2-dimensional $\delta$-function in the source term 
by using the Green's function method. 

In order to obtain the flat d'Alambertian, 
we consider the Schwarzschild metric in the isotropic coordinates, 
\begin{eqnarray}
ds^2 &=
- \frac{(2\rho-M)^2}{(2\rho+M)^2} dt^2+\left(1+\frac{M}{2\rho}\right)^{4} 
\left[ d\rho^2
+\rho^2\left(d\theta^2+\sin^2\theta d\phi^2\right) \right] \,. 
\end{eqnarray}
This radial coordinate is related to that of the Schwarzschild, r,
\begin{eqnarray}
\rho = \frac{r-M+(r^2-2Mr)^{1/2}}{2} \,. 
\end{eqnarray}
In the above coordinates, the Klein-Gordon equation with 
a point source reads 
\begin{eqnarray}
\fl 
\Biggl[
-{\frac { \left( 2\,\rho+M \right) ^{2}}
{ \left( 2\,\rho-M \right) ^{2}}}\partial_t^{2}
+ {\frac {16 {\rho}^{4}}{ \left( 2\,\rho+M \right) ^{4}}}\partial_{\rho}^{2}
+{\frac {128{\rho}^{5}}
{ \left( 2\,\rho - M \right)  \left( 2\,\rho+M \right) ^{5}}}\partial_\rho
\nonumber \\ \hspace{-10mm}
+{\frac {16 {\rho}^{2}}{ \left( 2\,\rho+M \right) ^{4}}}
\left(\partial_{\theta}^2 + \cot\theta \partial_{\theta} 
+\frac{1}{\sin^2 \theta}\partial_{\phi}^2 
\right)\Biggr] 
\times \psi(t,\rho,\theta,\phi) 
\nonumber \\ \hspace{-10mm}
= - q \int_{-\infty}^{\infty} d\tau 
\frac{64\rho^4\delta(t-t_z(\tau))\delta(\rho-\rho_z(\tau))\delta(\theta-\theta_z(\tau))
\delta(\phi-\phi_z(\tau))}
{(2\rho-M)(2\rho+M)^5 \sin\theta} 
 \,, 
\end{eqnarray}
where $\psi$ and $q$ denote a scalar field and 
a scalar charge, respectively. 
$z^{\alpha}(\tau)$ is the particle's trajectory 
with a proper time $\tau$. 
Here, we use the azimuthal mode decomposition, 
\begin{eqnarray}
\psi(t,r,\theta,\phi)=\sum_{m=-\infty}^{\infty}
\psi_m(t,\rho,\theta) \exp (im\phi) \,, 
\end{eqnarray}
and then obtain the wave equation for each $m$ mode as 
\begin{eqnarray}
\fl 
\Biggl[
-{\frac { \left( 2\,\rho+M \right) ^{2}}
{ \left( 2\,\rho-M \right) ^{2}}}\partial_t^{2}
+ {\frac {16 {\rho}^{4}}{ \left( 2\,\rho+M \right) ^{4}}}\partial_{\rho}^{2}
+{\frac {128{\rho}^{5}}
{ \left( 2\,\rho - M \right)  \left( 2\,\rho+M \right) ^{5}}}\partial_\rho
\nonumber \\ \hspace{-10mm}
+{\frac {16 {\rho}^{2}}{ \left( 2\,\rho+M \right) ^{4}}}
\left(\partial_{\theta}^2 + \cot\theta \partial_{\theta} 
-\frac{m^2}{\sin^2 \theta}
\right) \Biggr] \psi_m(t,\rho,\theta) 
\nonumber \\ \hspace{-10mm}
= - q \int_{-\infty}^{\infty} d\tau 
\frac{64\rho^4\delta(t-t_z(\tau))\delta(\rho-\rho_z(\tau))\delta(\theta-\theta_z(\tau))}
{(2\rho-M)(2\rho+M)^5 \sin\theta} \exp[-i m \phi_z(\tau)] \,.
\label{eq:psimEq}
\end{eqnarray}
For the above equation, 
we transform the field $\psi_m$ as 
\begin{eqnarray}
\psi_m(t,\rho,\theta) =2\,
\left[{\frac{\rho}
{\left( 2\,\rho+M \right)  \left( 2\,\rho-M \right) \sin \theta }} \right]^{1/2}
\, \chi_m(t,\rho,\theta) \,.
\label{eq:Scalartrans}
\end{eqnarray} 
Above, we have set 
the radial and angular factors to obtain 
the spatial part of the flat (2+1)-dimensional d'Alambertian. 
Then, $\chi_m$ satisfies the following equation
\begin{eqnarray}
\fl 
\Biggl\{
-\frac{1}{16}\frac{(2\rho+M)^6}{(2\rho-M)^2\rho^4}
\partial_t^2+\partial_{\rho}^2+\frac{1}{\rho} \partial_{\rho} 
\nonumber \\ \hspace{-10mm}
+\frac{1}{\rho^2}\left[ \partial_{\theta}^2 
-\frac{1}{\sin^2 \theta}
\left(m^2 - \frac{1}{4}\frac{(4\rho^2+M^2)^2-16\rho M \cos^2 \theta}
{(2\rho+M)^2(2\rho-M)^2} \right)
 \right] \Biggr\} \chi_m(t,\rho,\theta) 
\nonumber \\ \hspace{-10mm}
= - q \int_{-\infty}^{\infty} d\tau 
\frac{2\,\delta(t-t_z(\tau))\delta(\rho-\rho_z(\tau))\delta(\theta-\theta_z(\tau))}
{\left[{(2\rho-M)(2\rho+M)\rho \sin\theta}\right]^{1/2}} 
\exp[-im \phi_z(\tau)] \,.
\label{eq:medWE}
\end{eqnarray}
Note that from the relation $\psi_m \sim \chi_m / \rho^{1/2}$, 
the source term for $\chi_m$ becomes $\rho^{1/2}$ times 
that for $\psi_m$. 
This fact will be used in \sref{sec:COC}. 
Next, we define a new time coordinate 
\begin{eqnarray}
T = \int^t dt \, \frac{4\,(2\rho_z(t)-M)\rho_z(t)^2}{(2\rho_z(t)+M)^3} \,, 
\label{eq:Timetrans}
\end{eqnarray}
where $\rho_z$ is obtained by solving the geodesic equation 
and we use the fact that the proper time $\tau$ 
and the argument $t$ in $\rho_z$ 
are also related by the geodesic equation. 
From this, we derive an equation which can be divided into 
the (2+1)-dimensional d'Alambertian of the flat case $\Box^{(2+1)}$ and a remainder. 
\begin{eqnarray}
{\cal L}_m \, \chi_m(T,\rho,\theta) 
&= \left(\Box^{(2+1)} + {\cal L}_m^{rem} \right) \chi_m(T,\rho,\theta) 
\nonumber \\ 
&= S_m(T,\rho,\theta) \,,
\label{eq:formal}
\end{eqnarray}
where the differential operators are given by 
\begin{eqnarray}
\fl 
\Box^{(2+1)} = -\partial_T^2+\partial_\rho^2 
+\frac{1}{\rho}\partial_\rho + \frac{1}{\rho^2} \partial_{\theta}^2 \,, 
\nonumber \\ 
\fl 
{\cal L}_m^{rem} = 
\left[
1-\frac{(2\rho_z(T)-M)^2\rho_z(T)^4(2\rho+M)^6}{(2\rho_z(T)+M)^6(2\rho-M)^2\rho^4}
\right]\partial_T^2
\nonumber \\ \hspace{-10mm}
- \frac{2 (4\rho_z(T)-M) (2\rho_z(T)-M)(2\rho+M)^6 \rho_z(T)^3 M }
{ (2\rho_z(T)+M)^7(2\rho-M)^2\rho^4 }
\left(\frac{d\rho_z(T)}{dT}\right)
\partial_T
\nonumber \\ \hspace{-10mm}
-\frac{1}{\rho^2 \sin^2 \theta}
\left[m^2 - \frac{1}{4}\frac{(4\rho^2+M^2)^2-16\rho M \cos^2 \theta}
{(2\rho+M)^2(2\rho-M)^2} \right] 
\,, 
\label{eq:DFdecomp}
\end{eqnarray}
and the source term is shown to be 
\begin{eqnarray}
\fl 
S_m(T,\rho,\theta) = - q \int_{-\infty}^{\infty} d\tau 
\frac{2\,\delta(t(T)-t_z(\tau))\delta(\rho-\rho_z(\tau))\delta(\theta-\theta_z(\tau))}
{\left[{(2\rho-M)(2\rho+M)\rho \sin\theta}\right]^{1/2}} 
\exp[-im \phi_z(\tau)] \,. 
\label{eq:formalSource}
\end{eqnarray}
Here, it is noted that there is no $\rho$ and $\theta$ derivatives 
in ${\cal L}_m^{rem}$ of \eref{eq:DFdecomp} 
because all $\rho$ and $\theta$ derivatives are included in $\Box^{(2+1)}$. 

To remove the $\delta$-function in the source term, we set 
\begin{eqnarray}
\chi_m(T,\rho,\theta) &=& \chi_m^S(T,\rho,\theta) 
+ \chi_m^{rem}(T,\rho,\theta) \,,
\end{eqnarray} 
where we define the new functions, $\chi_m^S$ and $\chi_m^{rem}$ as
calculated from 
\begin{eqnarray}
\Box^{(2+1)} \chi_m^S(T,\rho,\theta) &= S_m(T,\rho,\theta) \,, 
\label{eq:chimS_eq}
\\ 
{\cal L}_m \,\chi_m^{rem}(T,\rho,\theta) 
&= -{\cal L}_m^{rem} \chi_m^S(T,\rho,\theta) 
\nonumber \\ & 
= S_m^{(eff)}(T,\rho,\theta) \,.
\label{eq:divisionEq}
\end{eqnarray}
The effective source $S_m^{(eff)}$ contains no $\delta$-function 
\footnote{This decomposition of $\chi_m$ does not have any physical-meaning, 
i.e., $\chi_m^S$ is not identified as the singular part to be removed 
in the self-force calculation. 
The physically and mathematically-meaningful singular part 
of the field which is called as the $S$-part, 
have been discussed in \cite{Detweiler:2002mi}. 
Recently, Vega and Detweiler \cite{Vega:2007mc} 
have discussed a new method to derive the retarded field 
by using a specific approximation to the $S$-part. 
}. 
Note that ${\cal L}_m^{rem}$ includes a second-order derivative. 
But, since the factor of $\partial_T^2$ is zero at the particle location, 
the singular behavior of the effective source $S_m^{(eff)}$ weakens. 

The derivation of the singular field $\chi_m^S$ 
can be performed through the Green's function 
of the (2+1)-dimensional flat case, 
\begin{eqnarray}
G(T,{\bf x};T',{\bf x'}) &= 
\frac{1}{2\pi}
\frac{1}{\left[
{(T-T')^2-|{\bf x} - {\bf x'}|^2}\right]^{1/2}}
\theta((T-T')-|{\bf x} - {\bf x'}|)
 \,,
\label{eq:Green}
\end{eqnarray}
where $\theta$ denotes the Heaviside step function, 
${\bf x}$ is a 2-dimensional spatial vector and 
the spatial distance is defined by 
\begin{eqnarray}
|{\bf x} - {\bf x'}|=
\left[\rho^2+{\rho'}^2-2\,\rho\,{\rho'}\cos(\theta-\theta')\right]^{1/2}
\,.
\end{eqnarray}
Some detail on the above Green's function 
is discussion in \sref{app:Gfn}. 
Next, $\chi_m^S$ is calculated by the following integral 
\begin{eqnarray}
\chi_m^S(T,\rho,\theta) &=& \int dT' \rho' d\rho' d\theta' \,
G(T,{\bf x};T',{\bf x'}) S_m(T',\rho',\theta') \,.
\label{eq:formalS}
\end{eqnarray} 
Here, it should be noted that the Green's function in the above 
multiple integration is given by analytically. 
Therefore, it is easy to perform the integration in 
\eref{eq:formalS} even for general orbits. 

As for the remaining field $\chi_m^{rem}$, we use numerical integrations. 
However, the effective source $S_m^{(eff)}$ is not amenable 
for direct numerical integrations due to its non-continuous behavior 
at the particle location. To explicitly obtain a source term 
 well behaved everywhere, we will apply the formulation discussed above 
to the case of a particle on a circular orbit.

%%%%%%%%%%%%%%%%%%%%%%%%%%%%%%%%%%%%%%%%%%%%%%%%%%%%%%%%%%%%%%%%%%%%%%
\section{Circular Orbit Case}\label{sec:COC}
%%%%%%%%%%%%%%%%%%%%%%%%%%%%%%%%%%%%%%%%%%%%%%%%%%%%%%%%%%%%%%%%%%%%%%

We consider a particle in a circular orbit given by
\begin{eqnarray}
z^{\alpha}(\tau) = \left\{u^t \tau,\,r_0,\,\frac{\pi}{2},\,u^{\phi} \tau \right\}
\,, 
\end{eqnarray}
where $r_0$ denotes the orbital radius 
in Schwarzschild coordinates.
The four velocity $u^{\alpha}$ is written by 
\begin{eqnarray}
u^t &= \left({\frac{r_0}{r_0-3M}}\right)^{1/2} \,,
\quad 
u^{\phi} &= \left[{\frac{M}{r_0^2(r_0-3M)}}\right]^{1/2}\,.
%\quad
%u^r &= 0\,,
%\quad
%u^\theta &= 0\,. 
\end{eqnarray}

The relationship between the new time coordinate $T$ 
and the Schwarzschild time $t$ can be obtained analytically 
\begin{eqnarray}
T = \frac{4\,(2\rho_0-M)\rho_0^2}{(2\rho_0+M)^3} \, t \,,
\end{eqnarray}
where 
\begin{eqnarray}
\rho_0 = \frac{r_0-M+\left({r_0^2-2Mr_0}\right)^{1/2}}{2} \,.
\end{eqnarray} 
Note that in general, for non circular orbits, 
we need a numerical integration to derive this relationship. 

In order to calculate the singular field, 
the Green's function in \eref{eq:Green} 
is rewritten in terms of $t$ as 
\begin{eqnarray}
G(T(t),{\bf x};T(t'),{\bf x'}) &= \frac{1}{2\pi}
\frac{1}{\left[{\frac{16\,(2\rho_0-M)^2\rho_0^4}{(2\rho_0+M)^6}(t-t')^2
-|{\bf x} - {\bf x'}|^2}\right]^{1/2}}
\nonumber \\ & \quad \times 
\theta
\left(\frac{4\,(2\rho_0-M)\rho_0^2}{(2\rho_0+M)^3}(t-t')-|{\bf x} - {\bf x'}|
\right) \,.
\end{eqnarray}
In the following, 
we will discuss the $m=0$ and $m \neq 0$ modes separately. 

The effective source for a final regularized function $\chi_m^{reg}$ 
must go like $\Or(\rho^{-2})$ for $\rho \rightarrow \infty$ 
in the case of the $m=0$ mode 
and $\Or(\rho^{-3/2})$ for the $m \neq 0$ mode. 
The reason is the following. 
First, for $\rho \rightarrow \infty$, 
there are only $\rho^{n/2}$ ($n$: integer) 
and $\ln \rho$ terms in the effective source of this paper. 
And, in practice, we will calculate the regularized function of 
the original field $\psi_m$ numerically. 
This means that the source term which we will use in the numerical calculation, 
becomes the factor $\sim 1/\rho^{1/2}$ times the source for $\chi_m^{reg}$. 
Thus, since we integrate the second order differential equation 
of \eref{eq:psimEq}, we need the above integrability conditions. 
In a similar manner, at the black hole horizon, 
i.e., $\rho \rightarrow M/2$, the source 
for $\psi_m^{reg}$ should vanish, 
i.e., the behavior of the source for $\chi_m^{reg}$ 
should be a power of $(\rho-M/2)$ greater than $1/2$ 
because the source for $\psi_m$ is 
the factor $\sim (\rho-M/2)^{-1/2}$ times the source for $\chi_m^{reg}$. 

%%%%%%%%%%%%%%%%%%%%%%%%%%%%%%%%%%%%%%
\subsection{The $m = 0$ modes}
%%%%%%%%%%%%%%%%%%%%%%%%%%%%%%%%%%%%%%

%%%%%%%%%%%%%%%%%%%
\subsubsection{Singular field}
%%%%%%%%%%%%%%%%%%%

In the Schwarzschild background, the source term 
in \eref{eq:formalSource} is time-independent, 
\begin{eqnarray}
S_0(t,r,\theta) &= - q 
\frac{2}{u^t\left[{(2\rho_0-M)(2\rho_0+M)\rho_0}\right]^{1/2}} 
\delta(\rho-\rho_0)\delta(\theta-\pi/2)
 \,. 
\end{eqnarray}
Therefore, the solution $\chi_0^S(t,r,\theta)$ is obtained from 
the 2-dimensional Poisson equation as 
\begin{eqnarray}
\chi_0^S(t,r,\theta) &= - \frac{q}{2\pi}
\frac{2\,\sqrt{\rho_0}}{u^t \left[{(2\rho_0+M) (2\rho_0-M)}\right]^{1/2} } 
\ln (|{\bf x} - {\bf x_z}|) \,, 
\label{eq:chi0S}
\end{eqnarray}
where the spatial difference is given by 
\begin{eqnarray}
|{\bf x} - {\bf x_z}|=\left({\rho^2+\rho_0^2-2\,\rho\, \rho_0 \sin \theta}
\right)^{1/2} \,,
\end{eqnarray} 
In practice, we must normalize the variables 
inside the logarithmic term in \eref{eq:chi0S}. 
We may take $|{\bf x} - {\bf x_z}|
\rightarrow |{\bf x} - {\bf x_z}|/M$, but we will ignore the factor $1/M$ 
in the following. 

%%%%%%%%%%%%%%%%%%%
\subsubsection{Local behavior}
%%%%%%%%%%%%%%%%%%%

\begin{figure}[ht]
\center
\epsfxsize=250pt
\epsfbox{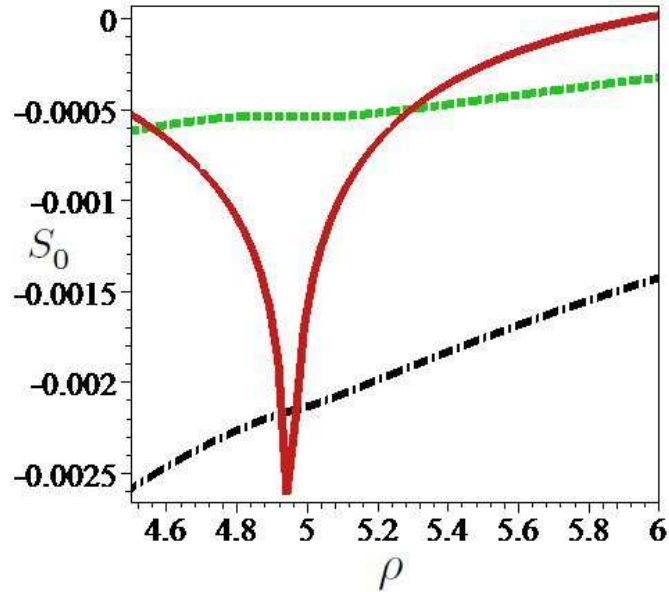}
\caption{
Plot for the $m=0$ mode of $S_m$ with respect to $\rho$ 
around the particle location. 
$S_0^{(eff)}$, $S_0^{reg,I}$, and $S_0^{reg,f}$
are shown by the solid red, dashed green, 
and dash-dotted black, respectively. 
The point particle is located at $\rho_0 \sim 4.95$ ($r_0=6$). 
}
\label{fig:middle0}
\end{figure}

The effective source for the $m=0$ mode, $S_0^{(eff)}$, 
is calculated by using the solution, $\chi_0^S$, as 
\begin{eqnarray}
S_0^{(eff)}(t,\rho,\theta) &= 
\frac{q}{8\pi}\frac{2\,\sqrt{\rho_0}}{u^t \left[{(2\rho_0+M) (2\rho_0-M)}\right]^{1/2} } 
\frac{1}{\rho^2 \sin^2 \theta}
\nonumber \\ & \quad \times 
\frac{(4\rho^2+M^2)^2-16\rho^2 M^2 \cos^2 \theta}
{(2\rho+M)^2(2\rho-M)^2} 
\ln (|{\bf x} - {\bf x_z}|)
\,.
\end{eqnarray}
This effective source is shown in \fref{fig:middle0} 
for the innermost stable circular orbit ($r_0=6M$) case, 
where $q=1$, $M=1$ and $\theta=\pi/2$
\footnote{When we show figures for the $m \neq 0$ mode, 
we also set $t=0$ and use the real part of sources.}. 
There is a singular behavior at the particle location. 
In order to perform the numerical integration with higher accuracy, 
it is convenient 
to regularize the source term to be at least $C^0$ at the particle location. 

In order to obtain the source $S_0^{reg,I}$ 
which is regular at the particle location, 
we choose the regularization function $\chi_0^{rem,S}$ as 
\begin{eqnarray}
\chi_0^{rem,S}(t,\rho,\theta) &= \frac{q}{16\pi}\,
{\frac { \rho_0^{15/2} \left(2\,\rho-M\right)^3 
\left( (4\rho^2+M^2)^2-16\rho^2 M^2 \cos^2 \theta \right) 
}
{u^t\, \left( 2\,\rho_0+M \right) ^{5/2} \left( 2\,\rho_0-M \right) ^{11/2}
\,{\rho}^{9}}}
\nonumber \\ & \quad \times 
|{\bf x} - {\bf x_z}|^2
\ln \left( |{\bf x} - {\bf x_z}| \right) 
\,.
\end{eqnarray}
This regularization function is chosen such that 
the additional source which arises from this function 
behaves well both 
for large $\rho$ and at the horizon. 
Here, the factors of the power of $1/\rho$ and $(2\rho-M)$ 
in the above equation give this good behavior. 
(We will discuss the $m \neq 0$ case by using the same treatment 
for the regularization function.) 

Although this regularization function 
(and the other regularization functions discussed later) 
is clearly not unique, the sum of the regularization functions 
and the regularized function, i.e., the (original) retarded function 
has the physical-meaning and is unique. 
This is because the source for the regularized function changes 
by differences of the regularization functions. 
Therefore, one may construct any appropriate source 
by which we can derive a regularized function. 

\begin{figure}[ht]
\center
\epsfxsize=250pt
\epsfbox{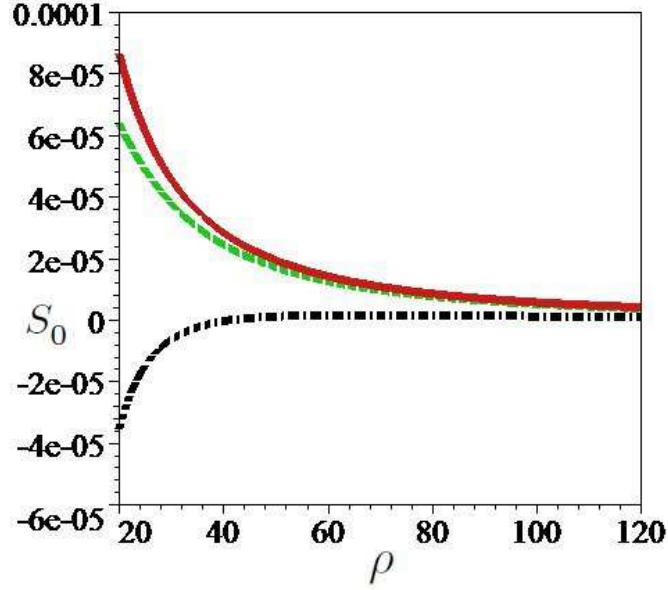}
\caption{
Plot for the $m=0$ mode of $S_m$ with respect to $\rho$ 
at large distance. 
$S_0^{(eff)}$, $S_0^{reg,I}$, and $S_0^{reg,f}$
are shown by the solid red, dashed green, 
and dash-dotted black, respectively. 
}
\label{fig:far0}
\end{figure}

\begin{figure}[ht]
\center
\epsfxsize=250pt
\epsfbox{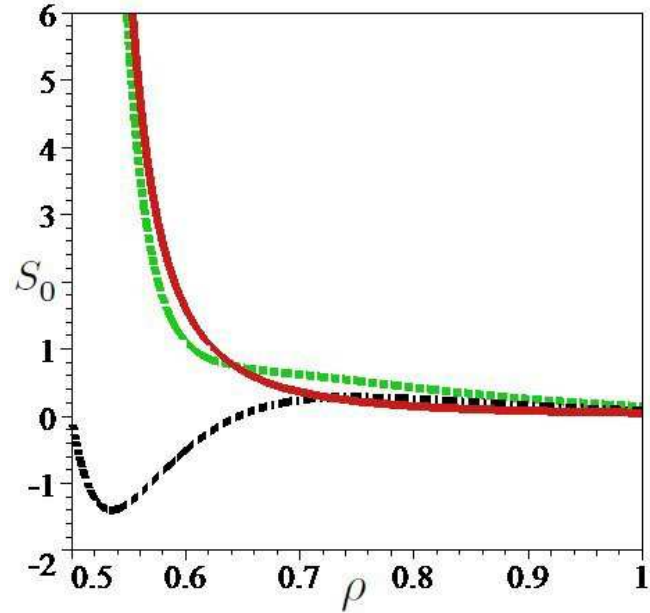}
\caption{
Plot for the $m=0$ mode of $S_m$ with respect to $\rho$ 
near the black hole horizon. 
$S_0^{(eff)}$, $S_0^{reg,I}$, and $S_0^{reg,f}$
are shown by the solid red, dashed green, 
and dash-dotted black, respectively. The location of the horizon 
is $\rho=0.5$. 
}
\label{fig:near0}
\end{figure}

Then, the regularized source at the particle location 
which is the source for $\chi_0^{rem}-\chi_0^{rem,S}$, 
is derived as 
\begin{eqnarray}
S_0^{reg,I}(t,\rho,\theta) &= S_0^{(eff)}(t,\rho,\theta) 
- {\cal L}_0 \chi_0^{rem,S}(t,\rho,\theta) 
\,.
\end{eqnarray}
We show the above source as 
the dashed green curve in \fref{fig:middle0}. 
This regular source $S_0^{reg,I}$ 
behaves as "$x \ln |x|$ for $x \rightarrow 0$", i.e., is 
$C^0$ around the particle location, 
and behaves as $\Or(1/\rho^2)$ for large $\rho$ 
(See \fref{fig:far0}.). 
However, it diverges as $\Or(1/(\rho-M/2)^2)$ 
at the horizon ($\rho=M/2$) (See \fref{fig:near0}.) . 
Therefore, we need one more regularization at the horizon. 

%%%%%%%%%%%%%%%%%%%
\subsubsection{Behavior at the horizon} 
%%%%%%%%%%%%%%%%%%%

Introducing the regularization function, 
\begin{eqnarray}
\fl 
\chi_0^h(t,\rho,\theta) = 
{\displaystyle \frac {q}{2 \,\pi}} \frac{\sqrt{\rho_0}}
{ u^t 
\left[{ (  2\,\rho_0 - M)\,(2\,\rho_0 + M)}\right]^{1/2}\,\rho ^{2}}
\nonumber \\ \hspace{-10mm} \times 
\Biggl[ {\displaystyle \frac {1}{4}} \,M^{2}\, \ln F(\theta) 
- \frac{M}{F(\theta)}(
- M^{2}\,\ln F(\theta) - 4\,\rho_0^{2}\,\ln F(\theta) 
+ 4\,\rho_0\,\sin \theta\,M\,\ln F(\theta) 
\nonumber \\ 
- M^{2} + 2\,\rho_0\,\sin \theta\,M)
\left(\rho  - {\displaystyle \frac {M}{2}} \right) 
- \Bigl(
- M^{4}\,\ln F(\theta) 
- 24\,M^{2}\,\rho_0^{2}\,\ln F(\theta) 
\nonumber \\ 
+ 8\,M^{3}\,\rho_0\,\sin \theta
\,\ln F(\theta) - 16\,\rho_0^{4}\,\ln F(\theta) 
+ 32\,\rho_0^{3}\,\sin \theta\,M\,\ln F(\theta) 
\nonumber \\ 
+ 16\,M^{2}\,\rho_0^{2}\,\cos^2 \theta\,\ln F(\theta) 
- 3\,M^{4} 
- 44\,M^{2}\,\rho_0^{2} + 20\,M^{3}\,\rho_0\,\sin \theta 
\nonumber \\ 
+ 32\,\rho_0^{3}\,\sin \theta\,M 
+ 24\,M^{2}\,\rho_0^{2}\,\cos^2 \theta \Bigr) 
\left(\rho  - {\displaystyle \frac {M}{2}} \right)^{2}
\left/ 
{\vrule height0.44em width0em depth0.44em} \right. \!  \! 
(M^{4} + 24\,M^{2}\,\rho_0^{2} 
\nonumber \\ 
- 8\,M^{3}\,\rho_0\,\sin \theta 
+ 16\,\rho_0^{4} - 32\,\rho_0^{3}\,\sin \theta\,M 
- 16\,M^{2}\,\rho_0^{2}\,\cos^2 \theta)
\Biggr]  \,,
\label{eq:reg0h}
\end{eqnarray}
where
\begin{eqnarray}
F(\theta) = \frac{1}{4}(M^{2} + 4\,\rho_0^{2} - 4\,\rho_0\,\sin \theta\,M) \,,
\label{eq:reg0hF}
\end{eqnarray}
we derive the following source for the regularized function, 
$\chi_0^{reg}=\chi_0^{rem}-\chi_0^{rem,S}-\chi_0^{h}$. 
\begin{eqnarray}
S_0^{reg,f}(t,\rho,\theta) = S_0^{reg,I}(t,\rho,\theta) 
- {\cal L}_0 \chi_0^{h}(t,\rho,\theta) \,.
\end{eqnarray}
Here, any bad behavior at infinity does not arise 
from the above regularization function 
because we have introduced the factor $1/\rho^2$ in \eref{eq:reg0h}. 
This regularization function have been derived 
by using the Taylor expansion around the horizon, $\rho=M/2$ 
(This treatment is also used to derive $\chi_m^{h}$ for 
$m \neq 0$ in \eref{eq:chimH}.) 
We find that the regularized source is $\Or(\rho-M/2)$ 
as shown by the dash-dotted black curve 
in \fref{fig:middle0}, \ref{fig:far0} and \ref{fig:near0} 
and remains of $\Or(\rho^{-2})$ for large $\rho$. 
This completes the results for the effective source term 
of the $m=0$ mode to be used in numerical calculations.

%%%%%%%%%%%%%%%%%%%%%%%%%%%%%%%%%%%%%%
\subsection{The $m \neq 0$ modes}
%%%%%%%%%%%%%%%%%%%%%%%%%%%%%%%%%%%%%%

%%%%%%%%%%%%%%%%%%%
\subsubsection{Singular field}
%%%%%%%%%%%%%%%%%%%

From \eref{eq:formalS}, 
the singular field for the $m \neq 0$ mode is derived as
\begin{eqnarray}
\fl 
\chi_m^S(t,r,\theta) = 
\frac{q}{2\pi} \int_{-\infty}^{\infty} d\tau 
\frac{8\,(2\rho_0-M)^{1/2} \rho_0^{5/2}}{(2\rho_0+M)^{7/2}} 
\frac{\exp({-im u^\phi \tau})}
{\left[{ \frac{16\,(2\rho_0-M)^2\rho_0^4}{(2\rho_0+M)^6} 
(t-u^t \tau)^2-|{\bf x} - {\bf x_z}|^2}\right]^{1/2}}
\nonumber \\ \qquad \times 
\theta
\left(\frac{4\,(2\rho_0-M)\rho_0^2}{(2\rho_0+M)^3}(t-u^t\tau)-|{\bf x} - {\bf x_z}|\right) 
\nonumber \\ \hspace{-5mm}
= 
\frac{q}{2\pi} \int_{-\infty}^{T_{ret}} dT_0
\, \frac{8\,(2\rho_0-M)^{1/2} \rho_0^{5/2}}{u^t(2\rho_0+M)^{7/2}} 
\nonumber \\  \qquad \times 
\frac{\exp[{-im (u^\phi/u^t) T_0}]}
{\left[{\frac{16\,(2\rho_0-M)^2\rho_0^4}{(2\rho_0+M)^6}(t-T_0)^2-|{\bf x} 
- {\bf x_z}|^2}\right]^{1/2}} \,, 
\end{eqnarray}
where the new variable for integration $T_0$, 
and the retarded time $T_{ret}$ are defined by 
\begin{eqnarray}
T_0 &= u^t \tau \,, 
\nonumber \\ 
T_{ret} &= t-\frac{(2\rho_0+M)^3}{4\,(2\rho_0-M)\rho_0^2}|{\bf x} - {\bf x_z}| \,. 
\end{eqnarray}
By introducing the following variable, 
\begin{eqnarray}
{\cal T} &= \frac{4\,(2\rho_0-M)\rho_0^2}{(2\rho_0+M)^3}
\,\frac{(t-T_0)}{|{\bf x} - {\bf x_z}|} \,, 
\end{eqnarray}
the above integration can be done as 
\begin{eqnarray}
\fl 
\chi_m^S(t,r,\theta) 
= 
\frac{q}{2\pi} \int_{1}^{\infty} d{\cal T} 
\frac{2\,\sqrt{\rho_0}}{u^t \left[{(2\rho_0+M) (2\rho_0-M)}\right]^{1/2} } 
\exp({-im \Omega t})
\nonumber \\ \times 
\frac{\exp\{im \Omega (2\rho_0+M)^3 |{\bf x} - {\bf x_z}|{\cal T}/ [4\,(2\rho_0-M)\rho_0^2] \}}
{\left({{\cal T}^2-1}\right)^{1/2}} 
\nonumber \\ \hspace{-5mm}
= \frac{i}{4}\,q\,\frac{2\,\sqrt{\rho_0}\,\exp({-im \Omega t}) }
{u^t \left[{(2\rho_0+M) (2\rho_0-M)}\right]^{1/2} } 
H_0^{(1)} 
\left(
\frac{(2\rho_0+M)^3}{4\,(2\rho_0-M)\rho_0^2} m \Omega |{\bf x} - {\bf x_z}|
\right) \,,
\end{eqnarray}
where the angular frequency is given by 
\begin{eqnarray}
\Omega &= \frac{u^\phi}{u^t} \,,
\end{eqnarray} 
and $H_0^{(1)}$ is the Hankel function of the first kind. 
The local behavior of the above solution 
near the particle location is 
\begin{eqnarray}
\fl 
\chi_m^S(t,r,\theta) \sim \chi_m^{SL}(t,r,\theta) 
\nonumber \\ \hspace{-6mm}
= - \frac{q}{2\pi}
\frac{2\,\sqrt{\rho_0}\,\exp({-im \Omega t}) }
{u^t \left[{(2\rho_0+M) (2\rho_0-M)}\right]^{1/2} } 
\ln 
\left(
\frac{(2\rho_0+M)^3}{4\,(2\rho_0-M)\rho_0^2} m \Omega |{\bf x} - {\bf x_z}|
\right) \,.
\end{eqnarray}

%%%%%%%%%%%%%%%%%%%
\subsubsection{Local behavior}
%%%%%%%%%%%%%%%%%%%

\begin{figure}[ht]
\center
\epsfxsize=250pt
\epsfbox{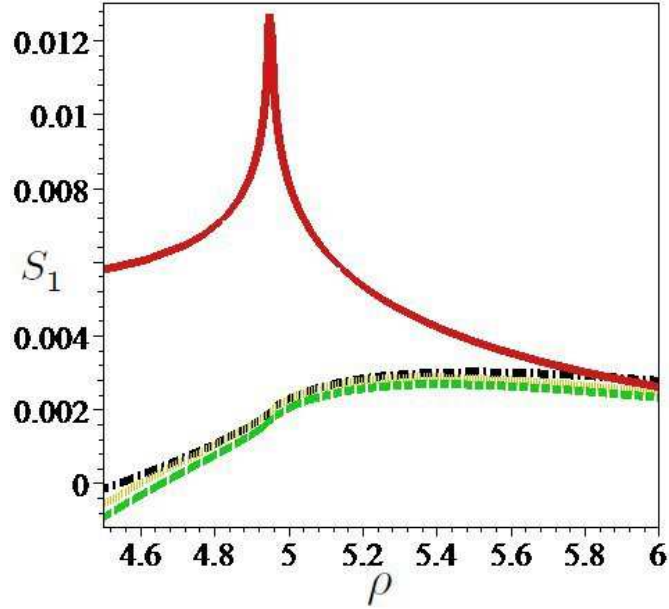}
\caption{
Plot for the $m=1$ mode of $S_m$ with respect to $\rho$ 
around the particle location. 
$S_1^{(eff)}$, $S_1^{reg,I}$, $S_1^{reg,h}$ and $S_1^{reg,f}$ 
are shown by the solid red, dashed green, dotted yellow 
and dash-dotted black, respectively. 
The point particle is located at $\rho_0 \sim 4.95$ ($r_0=6$). 
The $S_1^{reg,I}$, $S_1^{reg,h}$ and $S_1^{reg,f}$ curves 
have an almost same behavior in this region. 
}
\label{fig:middle1}
\end{figure}

When we write the singular field as 
\begin{eqnarray}
\chi_m^S = \chi_m^{SL} + {\hat \chi}_m^S \,, 
\end{eqnarray}
${\hat \chi}_m^S$ is finite at the particle location. 
Then, the effective source in \eref{eq:divisionEq} becomes 
\begin{eqnarray}
\fl 
S_m^{(eff)}(t,\rho,\theta) = - \frac{q}{2\pi}
\frac{2\,\sqrt{\rho_0}\,\exp({-im \Omega t})}
{u^t \left[{(2\rho_0+M) (2\rho_0-M)}\right]^{1/2} } 
\ln 
\left(
\frac{(2\rho_0+M)^3}{4\,(2\rho_0-M)\rho_0^2} m \Omega |{\bf x} - {\bf x_z}|
\right)
\nonumber \\ \times 
\Biggl(
-\frac{1}{\rho^2 \sin^2 \theta}
\left(m^2 - \frac{1}{4}\frac{(4\rho^2+M^2)^2-16\rho M \cos^2 \theta}
{(2\rho+M)^2(2\rho-M)^2} \right) 
\nonumber \\ \qquad 
-(m \Omega)^2\left(
\frac{(2\rho_0+M)^6}{(2\rho_0-M)^2\rho_0^4}
-\frac{(2\rho+M)^6}{(2\rho-M)^2\rho^4}
\right)
\Biggr)
- {\cal L}_m^{rem} {\hat \chi}_m^S(t,r,\theta)
\,,
\label{eq:cir_Smeff}
\end{eqnarray}
Note that the third line of the above right hand side 
is at least $C^0$ at the location of the particle. 
Therefore, $S_m^{(eff)}$ shown for the $m=1$ mode 
by the solid red curve in \fref{fig:middle1}, 
has a logarithmic divergence at the particle location. 
This behavior does not change for each $m$ mode. 

To remove the logarithmic divergence in the source, 
we introduce 
\begin{eqnarray}
\fl 
\chi_m^{rem,S}(t,\rho,\theta) = 
-\frac{q}{16{\pi }}\, |{\bf x} - {\bf x_z}|^2 
\ln 
\left(
\frac{(2\rho_0+M)^3}{4\,(2\rho_0-M)\rho_0^2} m \Omega |{\bf x} - {\bf x_z}|
\right)
\nonumber \\ \times 
\frac{{\rho_0}^{19/2} \left( 2\,\rho-M \right) ^{3}\, \exp({-im\Omega\,t}) }
{ u^t 
\left( 2\,{\rho_0}+M \right) ^{5/2} \left( 2\,{\rho_0}-M \right) ^{11/2} 
{\rho}^{11}\sin^2 \theta }
\left[
64\,{m}^{2}{\rho}^{4}-32\,{m}^{2}{\rho}^{2}{M}^{2}
\right. \nonumber \\ \qquad \left. 
+4\,{m}^{2}{M}^{4} 
+16\, \cos^2 \theta 
{\rho}^{2}{M}^{2}
-(4\rho^2+M^2)^2 
\right]
\,.
\label{eq:remS}
\end{eqnarray}
Using this regularization function, we obtain a source 
$S_m^{reg,I}$ for the function $\chi_m^{rem}-\chi_m^{rem,S}$
\begin{eqnarray}
S_m^{reg,I}(t,\rho,\theta) &=& S_m^{(eff)}(t,\rho,\theta) 
- {\cal L}_m \chi_m^{rem,S}(t,\rho,\theta) \,.
\end{eqnarray}
The local behavior of $S_m^{reg,I}$, 
which is shown by the dashed green curve 
in \fref{fig:middle1}, 
is of the form "$x \ln |x|$ for $x \rightarrow 0$", i.e., 
$C^0$ at the particle location. 

%%%%%%%%%%%%%%%%%%%
\subsubsection{Boundary behavior} 
%%%%%%%%%%%%%%%%%%%

\begin{figure}[ht]
\center
\epsfxsize=250pt
\epsfbox{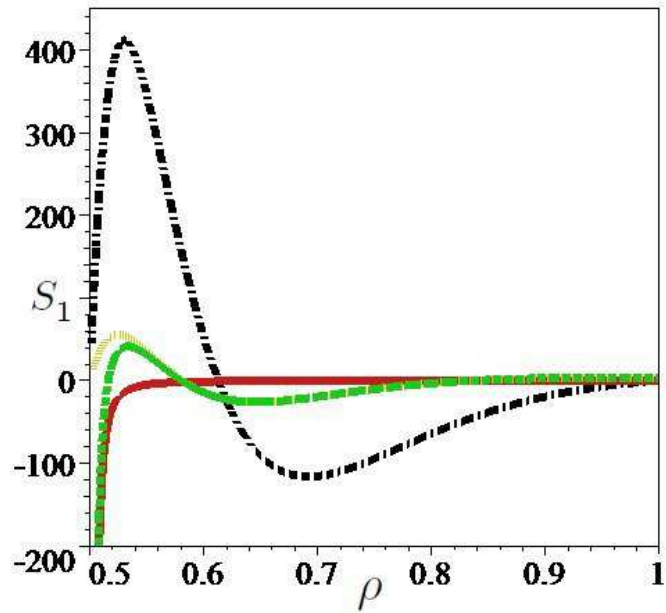}
\caption{
Plot for the $m=1$ mode of $S_m$ with respect to $\rho$ 
near the black hole horizon. 
$S_1^{(eff)}$, $S_1^{reg,I}$, $S_1^{reg,h}$ and $S_1^{reg,f}$ 
are shown by the solid red, dashed green, dotted yellow 
and dash-dotted black, respectively.
The $S_1^{reg,I}$ and $S_1^{reg,h}$ curves have an almost 
same behavior except near the horizon. 
The location of the horizon is $\rho=0.5$. 
}
\label{fig:near1}
\end{figure}

\begin{figure}[ht]
\center
\epsfxsize=250pt
\epsfbox{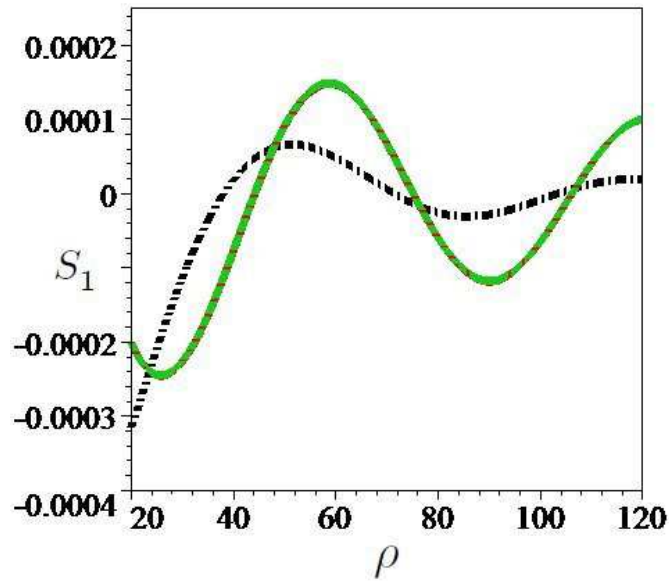}
\caption{
Plot for the $m=1$ mode of $S_m$ with respect to $\rho$ 
at large distance. 
$S_1^{(eff)}$, $S_1^{reg,I}$, $S_1^{reg,h}$ and $S_1^{reg,f}$ 
are shown by the solid red, dashed green, dotted yellow 
and dash-dotted black, respectively. 
The $S_1^{(eff)}$, $S_1^{reg,I}$ and $S_1^{reg,h}$ curves 
have an almost same behavior in this region. 
}
\label{fig:far1}
\end{figure}

We now focus on the behavior of the source term at the two boundaries, i.e., 
at the horizon of the large hole and spatial infinity. 
(See \fref{fig:near1} and \ref{fig:far1}.) 
To regularize the source at the boundaries, 
we note that the source contribution from $\chi_m^{rem,S}$ is well behaved. 
This means that the ill behaviors of the source 
arise from $\chi_m^S$. 
Therefore, it is convenient to use the asymptotic behavior 
of $\chi_m^S$ (and some correction factor) for regularization. 

For the regularization near the horizon, 
we use the regularization function $\chi_m^h$ 
given in \sref{app:chimh}. 
Then, the source 
for the function $\chi_m^{rem}-\chi_m^{rem,S}-\chi_m^h$ becomes 
\begin{eqnarray}
S_m^{reg,h}(t,\rho,\theta) &=& S_m^{reg,I}(t,\rho,\theta) 
- {\cal L}_m \chi_m^{h}(t,\rho,\theta) 
\,.
\end{eqnarray}
This $S_m^{reg,h}$ is shown by the dotted yellow curve in \fref{fig:near1} 
and vanishes as $\Or(\rho-M/2)$ at the horizon. 
But this source behaves as $\Or(\rho^{-1/2})$ for large $\rho$. 
(See \fref{fig:far1}.) To regularize it, 
we use the regularization function, 
\begin{eqnarray}
\fl 
\chi_m^{\infty}(t,\rho,\theta) = 
 - \left({\frac{2}{\pi }}\right)^{1/2}
\,i\,q\,\frac{\rho_0^{3/2}\,\exp( - i\,m\,\Omega \,t)}
{u^t\,(2\,\rho_0 + M)^{2}\,\left({m\,\Omega \,\rho }\right)^{1/2}\,\rho ^{7}} 
\,
(\rho ^{2} + \rho_0^{2} - 2\,\rho_0\,\rho 
\,\sin \theta)^{2} 
\nonumber \\ \hspace{-5mm} \times 
\left(\rho  - {\displaystyle \frac {M}{2}} \right)^{3} 
\,\exp
\Biggl[  \frac{i}{4}
\frac {(2\,\rho_0 + M)^{3}\,m\,\Omega \,
\left({\rho ^{2} + \rho_0^{2} - 2\,\rho_0\,\rho 
\,\sin \theta}\right)^{1/2}}{(2\,\rho_0 - M)\,\rho_0^{2}} 
- \frac {i\pi }{4}  \Biggr] 
\,. 
\end{eqnarray}
The final source for the regularized function 
$\chi_m^{reg}=\chi_m^{rem}-\chi_m^{rem,S}-\chi_m^h-\chi_m^{\infty}$ becomes 
\begin{eqnarray}
S_m^{reg,f}(t,\rho,\theta) &=& S_m^{reg,h}(t,\rho,\theta) 
- {\cal L}_m \chi_m^{\infty}(t,\rho,\theta) 
\,.
\end{eqnarray}
This $S_m^{reg,f}$ is shown by the dash-dotted black curve 
in \fref{fig:far1}, and behaves like $\Or(\rho^{-3/2})\, \times$ 
(an oscillation factor with respect to $\rho$) for large $\rho$. 
Using this effective source, 
we can calculate $\chi_m^{reg}$ by numerical calculations.  

%%%%%%%%%%%%%%%%%%%%%%%%%%%%%%%%%%%%%%%%%%%%%%%%%%%%%%%%%%%%%%%%%%%%%%
\section{Discussion}\label{sec:dis}
%%%%%%%%%%%%%%%%%%%%%%%%%%%%%%%%%%%%%%%%%%%%%%%%%%%%%%%%%%%%%%%%%%%%%%

In this paper, we have formulated how to derive a global effective source, 
in replacement of a two-dimensional Dirac's delta, 
for the (2+1)-dimensional Klein-Gordon differential equation 
on a black hole background 
by using a transformation of the scalar field in \eref{eq:Scalartrans} 
and a coordinate transformation with respect to time 
in \eref{eq:Timetrans}. 
Here, since we focus on the retarded field which is global, 
we do not use any local analysis and of the field, 
but have treated only the fields, 
i.e., $\chi_m^{S}$, $\chi_m^{rem,S}$, $\chi_m^{h}$ 
and $\chi_m^{\infty}$ defined globally. 
The above treatment of the regularization functions is 
the feature of this paper. 
On the other hand, Barack and Golbourn \cite{Barack:2007jh} 
have introduced a thin worldtube surrounding the worldline 
of a point particle, i.e., the local analysis 
to derive the retarded field. In this approach it is important
to obtain results that are insensitive to the choice of the
size of the world tube. On the other hand, our approach,
determining a global effective source, is straightforward to
use, once the regularization is done.

As the application, in the case of circular orbit, 
we obtained the regularized effective source 
$S_m^{reg,f}$ for the field $\chi_m^{reg}$. 
This source is $C^0$ at the location of the particle, 
and $\Or(\rho-M/2)$ near the horizon. 
The behavior at infinity is 
$\Or(\rho^{-2})$ for the $m=0$ mode and 
$\Or(\rho^{-3/2}) \, \times $ 
(an oscillation factor with respect to $\rho$) for 
the $m \neq 0$ modes, 
which allows straightforward numerical integration. 

In the case of general orbits, there is some additional difficulty. 
If it is possible to use the slow motion approximation 
and the eccentricity expansion for the bounded orbit cases, 
we can obtain $\chi_m^{S}$ analytically. 
Although we need to derive the singular field $\chi_m^{S}$ 
by numerical calculations in general, these 
include only the numerical integration 
for \eref{eq:Timetrans} and \eref{eq:formalS}. 
The regularization functions $\chi_m^h$ and $\chi_m^{\infty}$ 
for the field $\chi_m^{rem}$ are extracted 
from the asymptotic behavior directly. 
This is the same method used in the circular case. 
About the regularization functions $\chi_m^{rem,S}$, 
using \eref{eq:formalS} with the source 
which is evaluated from the asymptotic behavior of $S_m^{(eff)}$, 
we can derive $\chi_m^{rem,S}$ by a numerical integration 
because we need only the most singular part around the particle's location. 
Hence, there is no trouble 
to obtain the effective source for $\chi_m^{reg}$. 

When we consider the extension of this formulation 
to the Kerr background case, we can also extract 
a similar differential operator to that of \eref{eq:formal}. 
In practice, we have the (2+1)-dimensional d'Alambertian of 
the flat spacetime in this case. 
The mode decomposition only in the azimuthal direction 
has already been used, 
therefore, the same treatment discussed in this paper is applicable. 

Finally, in the case of gravitational perturbations, 
we have ten field equations for the linear perturbation 
in the Lorenz gauge. (See \cite{Barack:2005nr}.) 
We can also extract the (2+1)-dimensional d'Alambertian 
of the flat spacetime from them. 
The same treatment also holds to those equations. 
The method can also be used to deal with the Teukolsky differential 
equation \cite{Teukolsky:1973ha} with the corresponding corrections
for the presence of not only Dirac's delta but first and second derivatives
of it as source terms.

\ack

We would like to thank N.~Sago and  H.~Tagoshi for useful discussions. 
This is supported by JSPS for Research Abroad (HN), by the NSF through  
grants PHY-0722315,  PHY-0701566, PHY-0714388, and PHY-0722703, and
from grant NASA 07-ATFP07-0158.

\appendix

%%%%%%%%%%%%%%%%%%%%%%%%%%%%%%%%%%%%%%%%%%%%%%%%%%%%%%%%%%%%%%%%%%%%%%
\section{About Green's function}\label{app:Gfn} 
%%%%%%%%%%%%%%%%%%%%%%%%%%%%%%%%%%%%%%%%%%%%%%%%%%%%%%%%%%%%%%%%%%%%%%

To obtain $\chi_m^S$ from \eref{eq:chimS_eq}
we use the Green's function method. 
Here we consider the Green's function in the Cartesian coordinates, 
i.e., $x=r\cos \theta$ and $y=r \sin \theta$
\begin{eqnarray}
\left( -\partial_t^2 + \partial_x^2 + \partial_y^2 
\right) G(t,{\bf x};t',{\bf x'})  = -\delta(t-t')\delta^{(2)}({\bf x}-{\bf x'}) \,,
\end{eqnarray}
where $\delta^{(2)}$ denotes the 2-dimensional $\delta$-function. 
A solution of the above equation is 
usually calculated by using the Fourier transformation, 
\begin{eqnarray}
G(t,{\bf x};t',{\bf x'}) &=& 
\frac{1}{2\,\pi} \int_{-\infty}^{\infty} 
d\omega G_{\omega}({\bf x};t',{\bf x'}) \exp({-i \omega t}) \,.
\label{eq:Ffourie}
\end{eqnarray} 
In the frequency domain, the Green's function satisfies the equation, 
\begin{eqnarray}
\left( \omega^2 + \partial_x^2 + \partial_y^2 
\right) G_{\omega}({\bf x};t',{\bf x'}) 
= -\exp({i \omega t'})\delta^{(2)}({\bf x}-{\bf x'}) \,.
\end{eqnarray}
The Green's function must become a homogeneous solution of 
the above equation for ${\bf x} \neq {\bf x'}$. And, near the singularity, 
${\bf x} \rightarrow {\bf x'}$, the Green's function must behave as 
$G_{\omega} \sim -1/(2\pi) \ln |{\bf x} - {\bf x'}|$. 
Furthermore, we need to set a boundary condition. 
The out-going boundary condition is used here. 
From the above three conditions, the Green's function 
in the frequency domain is obtained as 
\begin{eqnarray}
G_{\omega}({\bf x};t',{\bf x'})  = \frac{i}{4} H_0^{(1)} 
(\omega |{\bf x} - {\bf x'}| ) \exp({i \omega t'}) \,,
\end{eqnarray}
where $H_0^{(1)}$ is the Hankel functions of the first kind. 

We go back to the time domain by using \eref{eq:Ffourie}. 
\begin{eqnarray}
G(t,{\bf x};t',{\bf x'}) 
&=& \frac{i}{8\pi} \int_{-\infty}^{\infty} H_0^{(1)} 
(\omega |{\bf x} - {\bf x'}| ) \exp[{-i \omega (t-t')}] \,.
\end{eqnarray}
Note that $H_0^{(1)}(-|\omega| |{\bf x} - {\bf x'}| )
= - H_0^{(2)}(|\omega| |{\bf x} - {\bf x'}| )$. 
Here, we use an integral representation of the Hankel functions 
\begin{eqnarray}
H_0^{(1)} (x) &=& \frac{-2i}{\pi} \int_1^{\infty} dt 
\frac{\exp({i x t})}{(t^2-1)^{1/2}} \,, 
\nonumber \\ 
H_0^{(2)} (x) &=& \frac{2i}{\pi} \int_1^{\infty} dt 
\frac{\exp({-i x t})}{(t^2-1)^{1/2}} \,, 
\end{eqnarray}
and then, the Green's function in the time domain is derived as 
\begin{eqnarray}
G(t,{\bf x};t',{\bf x'}) &=& \frac{1}{2\pi}
\frac{1}{\left[{(t-t')^2-|{\bf x} - {\bf x'}|^2}\right]^{1/2}}
\theta((t-t')-|{\bf x} - {\bf x'}|) \,.
\end{eqnarray}
From the above Heaviside step function, this Green's function has support 
not only on the light cone, but also inside the light cone. 
This is mentioned as a failure of the Huygens principle \cite{Kazinski:2002mp}. 
This feature of the Green's function has been discussed 
in Refs. \cite{Galtsov:2001iv,Kazinski:2002mp,Cardoso:2002pa}. 
The above Green's function is also derived from the direct integration 
of the (3+1)-dimensional Green's function 
with respect to the axial direction \cite{Galtsov:2001iv}.

%%%%%%%%%%%%%%%%%%%%%%%%%%%%%%%%%%%%%%%%%%%%%%%%%%%%%%%%%%%%%%%%%%%%%%
\section{Regularization function, $\chi_m^h$}\label{app:chimh} 
%%%%%%%%%%%%%%%%%%%%%%%%%%%%%%%%%%%%%%%%%%%%%%%%%%%%%%%%%%%%%%%%%%%%%%

The function $\chi_m^h$ for the regularization near the horizon 
is so long that we summarize it in this appendix. 
\begin{eqnarray}
\fl 
\chi_m^h(t,\rho,\theta) = 
{\displaystyle \frac {-1}{32}} \,i\,q\,
\frac{\sqrt{\rho_0}\,M^{4}}
{u^t 
\left[({2\,\rho_0 - M})\,({2\,\rho_0 + M})\right]^{1/2}\,\rho ^{4} }
\,\exp( - i\,m\,\Omega \,t)
\nonumber \\ 
\times \Biggl\{
C_0 (\rho,\,\theta)
H_0^{(1)} \left[
\frac{(2\rho_0+M)^3}{4\,(2\rho_0-M)\rho_0^2} m \Omega 
\sqrt{F(\theta)}\right] 
\nonumber \\ \qquad
+  C_1 (\rho,\,\theta)
H_1^{(1)} \left[
\frac{(2\rho_0+M)^3}{4\,(2\rho_0-M)\rho_0^2} m \Omega 
\sqrt{F(\theta)}\right] 
\nonumber \\ \qquad 
+C_2 (\rho,\,\theta)
H_2^{(1)} \left[
\frac{(2\rho_0+M)^3}{4\,(2\rho_0-M)\rho_0^2} m \Omega 
\sqrt{F(\theta)} \right] 
\Biggr\} \,,
\label{eq:chimH}
\end{eqnarray}
where
\begin{eqnarray}
\fl 
C_0 (\rho,\,\theta) =
1 
+ {\displaystyle \frac {8}{M}}\,\left(\rho  - {\displaystyle \frac {M}{2}} \right) 
- {\displaystyle \frac {1}{64}} \biggl[
64\,M^{4}\,m^{4}\,(2\,\rho_0 + M)^{6}(M^{4} 
+ 12\,M^{2}\,\rho_0^{2}
\,\cos^2 \theta 
\nonumber \\  
- 8\,M^{2}\,\rho_0^{2} 
 - 16\,M\,\cos^2 \theta\,\rho_0^{3}\,
\sin \theta 
- 16\,\rho_0^{4}\,
\cos^2 \theta + 16\,\rho_0^{4})\Omega ^{4} 
\nonumber \\  
+ M^{2}
\,m^{2}( 
29376\,M^{5}\,\rho_0^{5}\,\cos^2 \theta 
+ 
2448\,M^{7}\,\rho_0^{3}\,\cos^2 \theta 
- 
6274048\,\rho_0^{10} 
\nonumber \\  
 + 884\,M^{8}\,\rho_0^{2} + 17\,M^{10} 
- 3264\,M
^{6}\,\rho_0^{4}\,\cos^2 \theta\,
\sin \theta 
 - 43520\,M^{4}\,\cos^2 \theta\,\rho_0^{6}
\,\sin \theta 
\nonumber \\  
- 3128320\,\rho_0^{7}\,M^{3} 
+ 204\,M^{8}\,\rho_0^{2}\,\cos^2 \theta 
 - 52224\,M\,\rho_0^{9}\,\cos^2 \theta 
\nonumber \\  
- 52224\,M^{2}\,\rho_0^{8}\,\sin \theta\,
\cos^2 \theta 
+ 1088\,M^{7}\,\rho_0^{3} 
- 16320\,M^{5}\,\rho_0^{5}
\,\cos^2 \theta\,\sin \theta 
\nonumber \\  
 + 204\,\rho_0\,M^{9} 
- 17408\,\rho_0^{10}\,\cos^2 \theta 
 - 17408\,M\,\rho_0^{9}\,\sin \theta\,
\cos^2 \theta 
\nonumber \\  
- 65280\,M^{3}\,\rho_0^{7}\,
\sin \theta\,\cos^2 \theta 
 - 272\,M^{7}\,\rho_0^{3}\,\cos^2 \theta
\,\sin \theta + 377984\,\rho_0^{5}
\,M^{5} 
\nonumber \\  
- 6343680\,M^{2}\,\rho_0^{8}
\,\cos^2 \theta 
 - 1540224\,M^{4}\,\rho_0^{6}\,\cos^2 \theta 
+ 6343680\,M\,\rho_0^{9} 
\nonumber \\  
 + 6287104\,M^{3}\,\rho_0^{7}\,\cos^2 \theta 
+ 11968\,M^{6}\,\rho_0^{4}\,\cos^2 \theta 
+ 1629440\,M^{2}\,\rho_0^{8} 
\nonumber \\  
 + 377984\,M^{4}\,\rho_0^{6} - 102112\,M^{6}
\,\rho_0^{4})\Omega ^{2} 
- 13824\,\rho_0^{4}
\,( 2\,\rho_0 - M)^{2} (M^{4} 
\nonumber \\  
+ 16\,M^{2}\,\rho_0^{2}\,\cos^2 \theta 
- 8\,M^{2}\,\rho_0^{2} + 16\,\rho_0^{4}) \biggr]
\left(\rho  - {\displaystyle \frac {M}{2}} \right)^{2} \left/ {\vrule 
height0.44em width0em depth0.44em} \right. \!  \! 
\biggl[ ( 2\,\rho_0 - M)^{2}
\nonumber \\  \times 
(9 + 64\,M^{2}\,m^{2}\,\Omega ^{2}) \,\rho_0^{4}\,M^{2}
(M^{4} + 16\,M^{2}\,\rho_0^{2}\,\cos^2 \theta 
- 8\,M^{2}\,\rho_0^{2} + 16\,\rho_0^{4}) \biggr]   \,,
\nonumber \\ 
\fl 
C_1 (\rho,\,\theta) =
{\displaystyle 
\frac {1}{4}} (2\,\rho_0 + M)^{3}\,\Omega \,m(M^{9} - 18
\,\rho_0\,M^{8}\,\sin \theta 
 - 128\,M^{7}\,\rho_0^{2}\,\cos^2 \theta 
+ 144\,\rho_0^{2}\,M^{7} 
\nonumber \\  
+ 448\,\rho_0^{3}
\,M^{6}\,\sin \theta\,\cos^2 \theta 
 - 672\,M^{6}\,\rho_0^{3}\,\sin \theta
 + 768\,M^{5}\,\rho_0^{4}\,\cos^4 \theta
\nonumber \\  
- 2688\,M^{5}\,\rho_0^{4}\,\cos^2 \theta 
 + 2016\,M^{5}\,\rho_0^{4} - 4032\,\rho_0^{5}
\,M^{4}\,\sin \theta 
+ 5376\,\rho_0^{6}\,M^{3} 
\nonumber \\  
+ 3584\,\rho_0^{5}\,
M^{4}\,\sin \theta\,\cos^2 \theta 
 - 512\,\rho_0^{5}\,M^{4}\,\sin \theta
\,\cos^4 \theta
- 7168\,\rho_0^{6}\,M^{3}\,
\cos^2 \theta 
\nonumber \\  
 + 2048\,\rho_0^{6}\,M^{3}\,\cos^4 \theta
+ 3072\,\rho_0^{7}\,M^{2}\,\sin \theta\,
\cos^2 \theta  - 4608\,\rho_0^{7}\,M^{2}\,\sin \theta
\nonumber \\  
 + 2304\,\rho_0^{8}\,M - 2048\,\rho_0^{8}\,M\,
\cos^2 \theta 
 - 512\,\rho_0^{9}\,\sin \theta)
\left(\rho  - {\displaystyle \frac {M}{2}} \right) 
\nonumber \\  
\left/ {\vrule 
height0.56em width0em depth0.56em} \right. \!  \! 
\biggl[(M^{2} + 4\,
\rho_0^{2} - 4\,\rho_0\,\sin \theta\,M)^{9/2} 
\rho_0^{2}\,( - 2\,\rho_0 + M) \biggr]
\nonumber \\ 
+ 
\biggl[ 
64\,M^{2}\,m^{2}(2\,M^{9} + 288\,\rho_0^{2}\,M^{7} 
+ 7008\,\rho_0^{5}\,M^{4}\,\sin \theta\,
\cos^2 \theta + 10752\,\rho_0^{6}\,M^{3} 
\nonumber \\  
 - 14096\,\rho_0^{6}\,M^{3}\,\cos^2 \theta 
- 8064\,\rho_0^{5}\,M^{4}\,\sin \theta
 + 4032\,M^{5}\,\rho_0^{4} 
 - 1024\,\rho_0^{9}\,\sin \theta 
\nonumber \\  
- 9216
\,\rho_0^{7}\,M^{2}\,\sin \theta - 255\,M^{7}
\,\rho_0^{2}\,\cos^2 \theta 
 + 4608\,\rho_0^{8}\,M 
- 1344\,M^{6}\,\rho_0^{3}\,\sin \theta 
\nonumber \\  
+ 884\,\rho_0^{3}\,M^{6}
\,\sin \theta\,\cos^2 \theta 
 + 5952\,\rho_0^{7}\,M^{2}\,\sin \theta
\,\cos^2 \theta + 3904\,\rho_0^{6}\,M^{3}\,
\cos^4 \theta
\nonumber \\  
 - 4032\,\rho_0^{8}\,M\,\cos^2 \theta
 - 5316\,M^{5}\,\rho_0^{4}\,\cos^2 \theta 
 - 960\,\rho_0^{5}\,M^{4}\,\sin \theta
\,\cos^4 \theta
\nonumber \\  
+ 1488\,M^{5}\,\rho_0^{4}\,
\cos^4 \theta
- 36\,\rho_0\,M^{8}\,\sin \theta) 
\Omega ^{2} + 16\,M^{9} + 2368\,\rho_0^{2}\,M^{7}
\nonumber \\  
 + 40960\,\rho_0^{8}\,M + 33600\,\rho_0^{6}\,M^{3}
\,\cos^4 \theta 
 + 59744\,\rho_0^{5}\,M^{4}\,\sin \theta\,
\cos^2 \theta 
\nonumber \\  
- 80896\,\rho_0^{7}\,M^{2}
\,\sin \theta 
 - 11200\,M^{6}\,\rho_0^{3}\,\sin \theta 
- 121872\,\rho_0^{6}\,M^{3}\,\cos^2 \theta 
\nonumber \\  
 - 44804\,M^{5}\,\rho_0^{4}\,\cos^2 \theta 
+ 52032\,\rho_0^{7}\,M^{2}\,\sin \theta
\,\cos^2 \theta 
- 35776\,\rho_0^{8}\,M\,\cos^2 \theta 
\nonumber \\  
- 9216\,\rho_0^{9}\,\sin \theta + 93184\,
\rho_0^{6}\,M^{3} 
 - 8128\,\rho_0^{5}\,M^{4}\,\sin \theta
\,\cos^4 \theta
\nonumber \\  
+ 7348\,\rho_0^{3}\,M^{6}\,
\sin \theta\,\cos^2 \theta 
 - 68992\,\rho_0^{5}\,M^{4}\,\sin \theta 
+ 12496\,M^{5}\,\rho_0^{4}\,\cos^4 \theta
\nonumber \\  
 - 292\,\rho_0\,M^{8}\,\sin \theta 
 - 2095\,M^{7}\,\rho_0^{2}\,\cos^2 \theta 
+ 34048\,M^{5}\,\rho_0^{4} \biggr] m\,\Omega \,(2\,\rho_0 + M)^{3}
\nonumber \\  
\times \left(\rho  - {\displaystyle \frac {M}{2}} \right)^{2}
 \left/ {\vrule height0.56em width0em depth0.56em} \right. \! 
 \!  
\biggl[
( - 2\,\rho_0 + M)\,M\,\rho_0^{2}\,(M^{2} + 4
\,\rho_0^{2} - 4\,\rho_0\,\sin \theta\,M)^{9/2} 
\nonumber \\  
\times (9 + 64\,M^{2}\,m^{2}\,\Omega ^{2}) 
\biggr]   \,,
\nonumber \\
\fl 
C_2 (\rho,\,\theta) 
= {\displaystyle \frac {1}{64}} 
(M^{2} - 4\,\rho_0\,\sin \theta\,M + 4\,\rho_0^{2} 
- 4\,\cos^2 \theta\,\rho_0^{2})\,\Omega ^{2} \,m^{2}(2\,\rho_0 + M)^{6}
(64\,M^{2}\,m^{2}\,\Omega ^{2} \nonumber \\  
+ 1)
\,\left(\rho  - {\displaystyle \frac {M}{2}} \right)^{2}\,
\left/ 
{\vrule height0.44em width0em depth0.44em} \right. \!  \! 
\biggl[ 
F(\theta)\,( - 2\,\rho_0 + M)^{2}\,\rho_0^{4} 
(9 + 64\,M^{2}\,m^{2}\,\Omega ^{2}) \biggr]
\end{eqnarray}
In the above equations, $F(\theta)$ is the same as defined in \eref{eq:reg0hF}.

%%%%%%%%%%%%%%%%%


\section*{References}
\begin{thebibliography}{ederf}

\bibitem{LISA}
  K. Danzman {\it et al}, 
  {\it LISA -- Laser Interferometer Space Antenna, Pre-Phase A Report}, 
  Max-Planck-Institute fur Quantenoptic, Report MPQ 233 (1998). 

%\cite{Poisson:2004gg}
\bibitem{Poisson:2004gg}
  E.~Poisson,
  %``The Gravitational self-force,''
  {\it General relativity and gravitation. Proceedings of GR17}, 119 (2005) 
  [arXiv:gr-qc/0410127].
  %%CITATION = GR-QC/0410127;%%

%\cite{Hikida:2004hs}
\bibitem{Hikida:2004hs}
  W.~Hikida, H.~Nakano and M.~Sasaki,
  %``Self-force regularization in the Schwarzschild spacetime,''
  Class.\ Quant.\ Grav.\  {\bf 22}, S753 (2005)
  [arXiv:gr-qc/0411150].
  %%CITATION = CQGRD,22,S753;%%

\bibitem{CQGVol22No15}
  C.~O. {Lousto} (ed.), \emph{{Special issue: Gravitational Radiation from Binary
  Black Holes: Advances in the Perturbative Approach}}, 
  %vol. 
  Class. Quant. Grav. {\bf 22}, S543-S868, (2005).

%\cite{Nakano:2007hv}
\bibitem{Nakano:2007hv}
  H.~Nakano and C.~O.~Lousto,
  %``Regular second order perturbations of extreme mass ratio black hole
  %binaries,''
  arXiv:gr-qc/0701039.
  %%CITATION = GR-QC/0701039;%%

%\cite{Pretorius:2005gq}
\bibitem{Pretorius:2005gq}
  F.~Pretorius,
  %``Evolution of binary black hole spacetimes,''
  Phys.\ Rev.\ Lett.\  {\bf 95}, 121101 (2005)
  [arXiv:gr-qc/0507014].
  %%CITATION = PRLTA,95,121101;%%

%\cite{Campanelli:2005dd}
\bibitem{Campanelli:2005dd}
  M.~Campanelli, C.~O.~Lousto, P.~Marronetti and Y.~Zlochower,
  %``Accurate evolutions of orbiting black-hole binaries without excision,''
  Phys.\ Rev.\ Lett.\  {\bf 96}, 111101 (2006)
  [arXiv:gr-qc/0511048].
  %%CITATION = PRLTA,96,111101;%%

%\cite{Baker:2005vv}
\bibitem{Baker:2005vv}
  J.~G.~Baker, J.~Centrella, D.~I.~Choi, M.~Koppitz and J.~van Meter,
  %``Gravitational wave extraction from an inspiraling configuration of  merging
  %black holes,''
  Phys.\ Rev.\ Lett.\  {\bf 96}, 111102 (2006)
  [arXiv:gr-qc/0511103].
  %%CITATION = PRLTA,96,111102;%%

%\cite{Campanelli:2007ew}
\bibitem{Campanelli:2007ew}
  M.~Campanelli, C.~O.~Lousto, Y.~Zlochower and D.~Merritt,
  %``Large Merger Recoils and Spin Flips From Generic Black-Hole Binaries,''
  Astrophys.\ J.\  {\bf 659}, L5 (2007)
  [arXiv:gr-qc/0701164].
  %%CITATION = ASJOA,659,L5;%%

%\cite{Campanelli:2007cga}
\bibitem{Campanelli:2007cga}
  M.~Campanelli, C.~O.~Lousto, Y.~Zlochower and D.~Merritt,
  %``Maximum gravitational recoil,''
  Phys.\ Rev.\ Lett.\  {\bf 98}, 231102 (2007)
  [arXiv:gr-qc/0702133].
  %%CITATION = PRLTA,98,231102;%%

%\cite{Lousto:2007db}
\bibitem{Lousto:2007db}
  C.~O.~Lousto and Y.~Zlochower,
  %``Further insight into gravitational recoil,''
  Phys.\ Rev.\  D {\bf 77}, 044028 (2008)
  [arXiv:0708.4048 [gr-qc]].
  %%CITATION = PHRVA,D77,044028;%%

%\cite{Gonzalez:2006md}
\bibitem{Gonzalez:2006md}
  J.~A.~Gonzalez, U.~Sperhake, B.~Bruegmann, M.~Hannam and S.~Husa,
  %``Total recoil: the maximum kick from nonspinning black-hole binary
  %inspiral,''
  Phys.\ Rev.\ Lett.\  {\bf 98}, 091101 (2007)
  [arXiv:gr-qc/0610154].
  %%CITATION = PRLTA,98,091101;%%

%\cite{Mino:1996nk}
\bibitem{Mino:1996nk}
  Y.~Mino, M.~Sasaki and T.~Tanaka,
  %``Gravitational radiation reaction to a particle motion,''
  Phys.\ Rev.\  D {\bf 55}, 3457 (1997)
  [arXiv:gr-qc/9606018].
  %%CITATION = PHRVA,D55,3457;%%

%\cite{Quinn:1996am}
\bibitem{Quinn:1996am}
  T.~C.~Quinn and R.~M.~Wald,
  %``An axiomatic approach to electromagnetic and gravitational radiation
  %reaction of particles in curved spacetime,''
  Phys.\ Rev.\  D {\bf 56}, 3381 (1997)
  [arXiv:gr-qc/9610053].
  %%CITATION = PHRVA,D56,3381;%%

%\cite{Detweiler:2002mi}
\bibitem{Detweiler:2002mi}
  S.~Detweiler and B.~F.~Whiting,
  %``Self-force via a Green's function decomposition,''
  Phys.\ Rev.\  D {\bf 67}, 024025 (2003)
  [arXiv:gr-qc/0202086].
  %%CITATION = PHRVA,D67,024025;%%

\bibitem{Lousto99b}
  C.~O.~Lousto,
  %``Pragmatic approach to gravitational radiation reaction in binary black
  %holes,''
  Phys.\ Rev.\ Lett.\  {\bf 84}, 5251 (2000)
  [arXiv:gr-qc/9912017].
  %%CITATION = PRLTA,84,5251;%%

%\cite{Barack:2001gx}
\bibitem{Barack:2001gx}
  L.~Barack, Y.~Mino, H.~Nakano, A.~Ori and M.~Sasaki,
  %``Calculating the gravitational self force in Schwarzschild spacetime,''
  Phys.\ Rev.\ Lett.\  {\bf 88}, 091101 (2002)
  [arXiv:gr-qc/0111001].
  %%CITATION = PRLTA,88,091101;%%

%\cite{Mino:2001mq}
\bibitem{Mino:2001mq}
  Y.~Mino, H.~Nakano and M.~Sasaki,
  %``Covariant self-force regularization of a particle orbiting a  Schwarzschild
  %black hole: Mode decomposition regularization,''
  Prog.\ Theor.\ Phys.\  {\bf 108}, 1039 (2003)
  [arXiv:gr-qc/0111074].
  %%CITATION = PTPKA,108,1039;%%

%\cite{Barack:2002bt}
\bibitem{Barack:2002bt}
  L.~Barack and A.~Ori,
  %``Regularization parameters for the self force in Schwarzschild spacetime:
  %II. gravitational and electromagnetic cases,''
  Phys.\ Rev.\  D {\bf 67}, 024029 (2003)
  [arXiv:gr-qc/0209072].
  %%CITATION = PHRVA,D67,024029;%%

%\cite{Barack:2002ku}
\bibitem{Barack:2002ku}
  L.~Barack and C.~O.~Lousto,
  %``Computing the gravitational self-force on a compact object plunging into a
  %Schwarzschild black hole,''
  Phys.\ Rev.\  D {\bf 66}, 061502 (2002)
  [arXiv:gr-qc/0205043].
  %%CITATION = PHRVA,D66,061502;%%

%\cite{Barack:2005nr}
\bibitem{Barack:2005nr}
  L.~Barack and C.~O.~Lousto,
  %``Perturbations of Schwarzschild black holes in the Lorenz gauge: Formulation
  %and numerical implementation,''
  Phys.\ Rev.\  D {\bf 72}, 104026 (2005)
  [arXiv:gr-qc/0510019].
  %%CITATION = PHRVA,D72,104026;%%

%\cite{Nakano:2003he}
\bibitem{Nakano:2003he}
  H.~Nakano, N.~Sago and M.~Sasaki,
  %``Gauge problem in the gravitational self-force: First post-Newtonian  force
  %under Regge-Wheeler gauge,''
  Phys.\ Rev.\  D {\bf 68}, 124003 (2003)
  [arXiv:gr-qc/0308027].
  %%CITATION = PHRVA,D68,124003;%%

\bibitem{Lousto97b}
  C.~O.~Lousto and R.~H.~Price,
  %``Understanding initial data for black hole collisions,''
  Phys.\ Rev.\  D {\bf 56}, 6439 (1997)
  [arXiv:gr-qc/9705071].
  %%CITATION = PHRVA,D56,6439;%%

\bibitem{Lousto:2005qh}
  C.~O.~Lousto,
  %``A time-domain fourth-order-convergent numerical algorithm to integrate
  %black hole perturbations in the extreme-mass-ratio limit,''
  Class.\ Quant.\ Grav.\  {\bf 22}, S543 (2005)
  [arXiv:gr-qc/0503001].
  %%CITATION = CQGRD,22,S543;%%

%\cite{Barack:2007jh}
\bibitem{Barack:2007jh}
  L.~Barack and D.~A.~Golbourn,
  %``Scalar-field perturbations from a particle orbiting a black hole using
  %numerical evolution in 2+1 dimensions,''
  Phys.\ Rev.\  D {\bf 76}, 044020 (2007)
  [arXiv:0705.3620 [gr-qc]].
  %%CITATION = PHRVA,D76,044020;%%

%\cite{LopezAleman:2003ik}
\bibitem{LopezAleman:2003ik}
  R.~Lopez-Aleman, G.~Khanna and J.~Pullin,
  %``Perturbative evolution of particle orbits around Kerr black holes: Time
  %domain calculation,''
  Class.\ Quant.\ Grav.\  {\bf 20}, 3259 (2003)
  [arXiv:gr-qc/0303054].
  %%CITATION = CQGRD,20,3259;%%

%\cite{Pound:2007th}
\bibitem{Pound:2007th}
  A.~Pound and E.~Poisson,
  %``Osculating orbits in Schwarzschild spacetime, with an application to
  %extreme mass-ratio inspirals,''
  Phys.\ Rev.\  D {\bf 77}, 044013 (2008)
  [arXiv:0708.3033 [gr-qc]].
  %%CITATION = PHRVA,D77,044013;%%

%\cite{Mino:2003yg}
\bibitem{Mino:2003yg}
  Y.~Mino,
  %``Perturbative Approach to an orbital evolution around a Supermassive black
  %hole,''
  Phys.\ Rev.\  D {\bf 67}, 084027 (2003)
  [arXiv:gr-qc/0302075].
  %%CITATION = PHRVA,D67,084027;%%

%\cite{Barack:2007we}
\bibitem{Barack:2007we}
  L.~Barack, D.~A.~Golbourn and N.~Sago,
  %``m-Mode Regularization Scheme for the Self Force in Kerr Spacetime,''
  Phys.\ Rev.\  D {\bf 76}, 124036 (2007)
  [arXiv:0709.4588 [gr-qc]].
  %%CITATION = PHRVA,D76,124036;%%

%\cite{Vega:2007mc}
\bibitem{Vega:2007mc}
  I.~Vega and S.~Detweiler,
  %``Regularization of fields for self-force problems in curved spacetime:
  %foundations and a time-domain application,''
  arXiv:0712.4405 [gr-qc].
  %%CITATION = ARXIV:0712.4405;%%

%\cite{Teukolsky:1973ha}
\bibitem{Teukolsky:1973ha}
  S.~A.~Teukolsky,
  %``Perturbations of a rotating black hole. 1. Fundamental equations for
  %gravitational electromagnetic, and neutrino field perturbations,''
  Astrophys.\ J.\  {\bf 185}, 635 (1973).
  %%CITATION = ASJOA,185,635;%%

%\cite{Kazinski:2002mp}
\bibitem{Kazinski:2002mp}
  P.~O.~Kazinski, S.~L.~Lyakhovich and A.~A.~Sharapov,
  %``Radiation reaction and renormalization in classical electrodynamics of
  %point particle in any dimension,''
  Phys.\ Rev.\ D {\bf 66}, 025017 (2002)
  [arXiv:hep-th/0201046].
  %%CITATION = HEP-TH 0201046;%%

%\cite{Galtsov:2001iv}
\bibitem{Galtsov:2001iv}
  D.~V.~Galtsov,
  %``Radiation reaction in various dimensions,''
  Phys.\ Rev.\ D {\bf 66}, 025016 (2002)
  [arXiv:hep-th/0112110].
  %%CITATION = HEP-TH 0112110;%%

%\cite{Cardoso:2002pa}
\bibitem{Cardoso:2002pa}
  V.~Cardoso, O.~J.~C.~Dias and J.~P.~S.~Lemos,
  %``Gravitational radiation in D-dimensional spacetimes,''
  Phys.\ Rev.\ D {\bf 67}, 064026 (2003)
  [arXiv:hep-th/0212168].
  %%CITATION = HEP-TH 0212168;%%

\end{thebibliography}
\end{document}